\documentclass[11pt]{article}
\usepackage{genres}
\usepackage[letterpaper,left=1.0in,right=1.0in,top=1.0in,bottom=1.0in]{geometry}               
\usepackage{graphicx}
\usepackage{extarrows}
\usepackage[utf8]{inputenc} 

\usepackage{amssymb}

\newtheorem{theorem}{Theorem}

\newcommand{\esp}[1]{\mathrm{E}\left [ #1\right ]}
\newcommand{\var}[1]{\mathrm{Var}\left [ #1\right ]}

\newcommand{\url}[1]{\texttt{#1}}

\newenvironment{skproof}{\noindent\textit{Sketch of proof.}}{~\hfill$\Box$}

\begin{document}
\thispagestyle{empty}

\begin{center}

%PUT THE TITLE HERE
 {\fontsize{14}{16} \textbf{
Integrating heterogeneous knowledges for understanding biological behaviors:  a probabilistic approach
} }\\
%PUT THE AUTHORS HERE
\fontsize{12}{14}
\vspace{6pt}
J\'{e}r\'{e}mie Bourdon$^{1*}$, Damien Eveillard$^{1}$, Samuel Gabillard$^{1}$, Theo Merle$^{2}$\\
\vspace{6pt}
1. LINA - CNRS and University of Nantes, Nantes, France\\
2. Ecole Normale Sup\'{e}rieure (ENS) - Cachan (Bretagne),  Bruz, France\\
$^*$email: jeremie.bourdon@univ-nantes.fr

\end{center}
\fontsize{12}{14}

\begin{abstract}

Despite recent molecular technique improvements, biological knowledge remains incomplete. Reasoning on living systems hence implies to integrate heterogeneous and partial informations. Although current investigations successfully focus on qualitative behaviors of macromolecular networks, others approaches show partial quantitative informations like protein concentration variations over times. We consider that both informations, qualitative and quantitative, have to be combined into a modeling method to provide a better understanding of the biological system. We propose here such a method using a probabilistic-like approach. After its exhaustive description, we illustrate its advantages by modeling the carbon starvation response in \textit{Escherichia coli}. In this purpose, we build an original qualitative model based on available observations. After the formal verification of its qualitative properties, the probabilistic model  shows quantitative results corresponding to biological expectations which confirm the interest of our probabilistic approach. 
%
%Despite recent molecular technique improvements, biological knowledge remains incomplete. Reasoning on living systems hence implies to integrate  heterogeneous and partial informations. Although current investigations successfully focus on qualitative behaviors of macromolecular networks, others approaches show partial quantitative informations like protein concentration variations over times. Both informations, qualitative and quantitative, have to be combined into a modeling method to provide a better understanding of the biological system. In this purpose, we propose here a new theoretical probabilistic-like approach. After its exhaustive description, we illustrate our method by modeling the carbon starvation response in \textit{Escherichia coli}. We build a qualitative model based on available observations. After a formal verification of their qualitative properties, further biological investigations confirm the interest of our probabilistic approach.

% besoin d'intégrer des informations partielles pour comprendre un systèmes vivant par définition incomplet. On propose pour cela une nouvelle approche probabiliste qui s'appuie sur toll based theory. Après une présentation de la méthode, nous en illustrerons les avantages sur un modele de réseaux de gènes de E coli soumis à une carence de source carbonées.   
\end{abstract}

\section{Introduction}
\begin{figure}[t]
\begin{center}
\includegraphics[scale=.6]{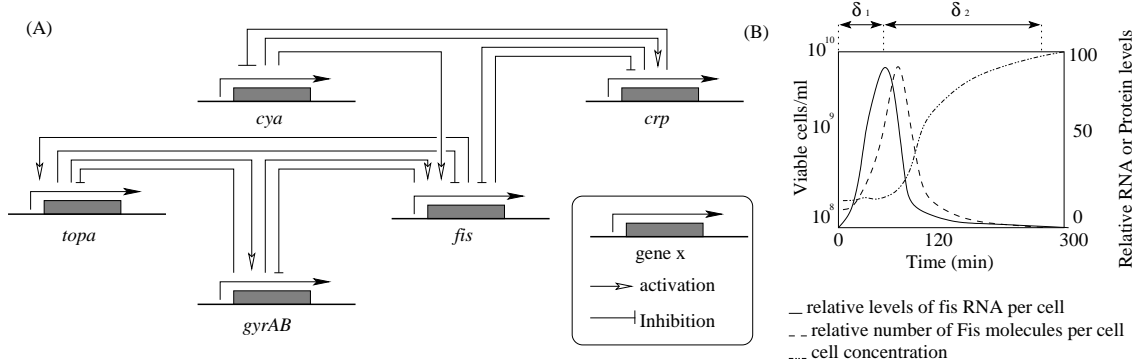}
\caption{Biological informations concerning \textit{Escherichia coli} carbon starvation system. (A) represents interactions between genes involved in the regulatory network (adapted from \cite{Ropers:2006aa}). (B) shows quantitative variations of macromolecules of interest (based on \cite{Ball:1992aa}). Note
%(i)
 the linear relationship between \textit{fis} RNA and  Fis protein productions.
% , as well as (ii) the ability to emphasize the delay associated with the molecular variation ($\delta_1$ or $\delta_2$ delays associated respectively with Fis concentration increase and decrease
}\vspace{1 mm}
\label{motivation}
\end{center}
\end{figure}

The last decade has seen great successes in macromolecular network modeling. In particular, qualitative methods appear today as well-adapted for reasoning on biological systems, despite the current lack of quantitative informations \cite{deJong:2002aa}. Thus most of interesting and investigated knowledges concern local informations such as gene-gene or gene-protein interactions. They allow to build networks like on Figure~\ref{motivation}~(A), that model the global qualitative behavior of a biological system. However, other experiments illustrated Figure~\ref{motivation}~(B) give insights about various partial quantitative knowledges. They emphasize both molecular concentration variations and time-series. These two related kinds of partial quantitative information, \textit{i.e.,} time and concentration, are well studied by other experiments \cite{Wolfe:2005aa}  and reflect as well the overall system behavior. Both informations, qualitative and quantitative, have hence to be combined into a modeling method for giving a better understanding of the biological system.
Due to the lack of quantitative informations, we propose a modeling approach that (i) spreads partial local informations through the qualitative network and (ii) gives insights about global behaviors.  Probabilistic approaches are well adapted  for bringing complementary quantitative or semi quantitative knowledges into a qualitative model. Among them, we suggest an original toll based approach that predicts various molecular productions combining both qualitative and partial quantitative knowledges. After an overview of our probabilistic approach (Sec.~2), we propose here to apply it on gene regulatory model of the carbon starvation response in \textit{Escherichia coli}. In this purpose, we (Sec.~3.1) build a model based on a novel qualitative abstraction, validate its behavior using a formal verification approach, which (Sec.~3.2) allows us to accurately apply our probabilistic method. Such a protocol emphasizes several biological insights of interest.

\section{Method}
\subsection{Biological system formalization}\label{formalization}
We consider biological networks as graphs that show transitions between various components of the system. Each transition is related to variations of characteristic quantities of the system and produces its own impact on the whole system behavior. In a gene regulatory network, a qualitative graph arrow is associated with a production or consumption of the corresponding protein. 
 
In order to abstract qualitative biological behaviors, we represent a gene regulatory network by a qualitative graph where each state stands for a qualitative variation of a gene activity. We focus on the macromolecular transformation derivative, which is more tractable to model detailed macromolecular concentration variations.  As illustration, following interactions describe the fact that $(i)$ gene $x$ activates gene $y$ and $(ii)$ $x$ represses $y$:
$$(i) \quad x \longrightarrow y{^+} \quad \quad (ii)  \quad x \longrightarrow y{^-}$$
Such a representation implies that  gene $x$ produces protein $X$ that activates gene $y$. Thus $(i)$ and $(ii)$ represent respectively an overall increase of $Y$ protein production and an overall decrease of $Y$. Note that such an abstraction neglects post-transcriptional regulations which is particularly unappropriated for modeling eukaryote gene regulatory network. 

This biological abstraction allows us to model various qualitative interactions. Considering that a gene $x$ activity is summarized by two qualitative states $x^+$ and $x^-$, $y$ activation by $x$ might be described by the set of rules and its corresponding transitions:
$$  \{x^+ \implies y{^+}\} \wedge  \{x^- \implies y{^+}\} $$
A peak of gene $x$ activity that activates $y$ is represented by:
$$  \{x^+ \implies y{^+}\} \wedge  \{x^- \implies \emptyset \} $$
A minimal activity of gene $x$ that activates $y$ is symbolized by:
$$  \{x^+ \implies \emptyset \} \wedge  \{x^- \implies y{^-} \} $$
Gene $y$ repressions by a gene $x$ activity are modeled using similar rules that imply transitions toward $y^-$. Such an abstraction gives the opportunity to focus on qualitative behaviors.  Reasoning on quantities associated with qualitative rules allows us to emphasize quantitative states of the system despite concurrent qualitative rules. 

\subsection{Graph model and quantities}

We make the assumption that the biological system is associated with several \emph{quantities} $q_1,\dots,q_k$ that represent the current state of the system. For illustration, these quantities represent protein concentrations, or other non trivial quantities such as the number of times a particular pathway is taken by the living system. Studying the behavior of biological systems hence consists in understanding the evolution of these quantities. Note here that such quantities may not be experimentally measurable. 
Since the last decade, biological behaviors have been often described by qualitative graphs that abstract different component variations within the system. In our model, we consider that each transition of this qualitative graph implies a potential variation on each quantity. Here we propose a method that focusses on these quantities.  

We consider two types of quantities. Some quantity variations are  additive whereas others are multiplicative. (i) Each transition from $i$ to $j$ is associated with  a real number $\delta_{(i,j)}$, the quantity $q$ is additive  if the quantity $q = x$ before the transition becomes  $x+\delta_{(i,j)}$ after the transition.
   (ii) Each transition from $i$ to $j$ is associated with  a strictly positive real number $\lambda_{(i,j)}$, the quantity $q$ is multiplicative  if the quantity $q = x$ before the transition becomes   $x\lambda_{(i,j)}$ after the transition. 
Each quantity $q$ is thus associated with a matrix $C_q$ in which the element at position $(i,j)$ is the contribution of the transition from $i$ to $j$. We are looking for understanding the  typical behavior of given additive or multiplicative quantities after a given time. These behaviors are controlled by an accumulation of small contributions.
For illustration, we consider the following graph.

\begin{minipage}{5cm}
$$
\includegraphics[width=4cm]{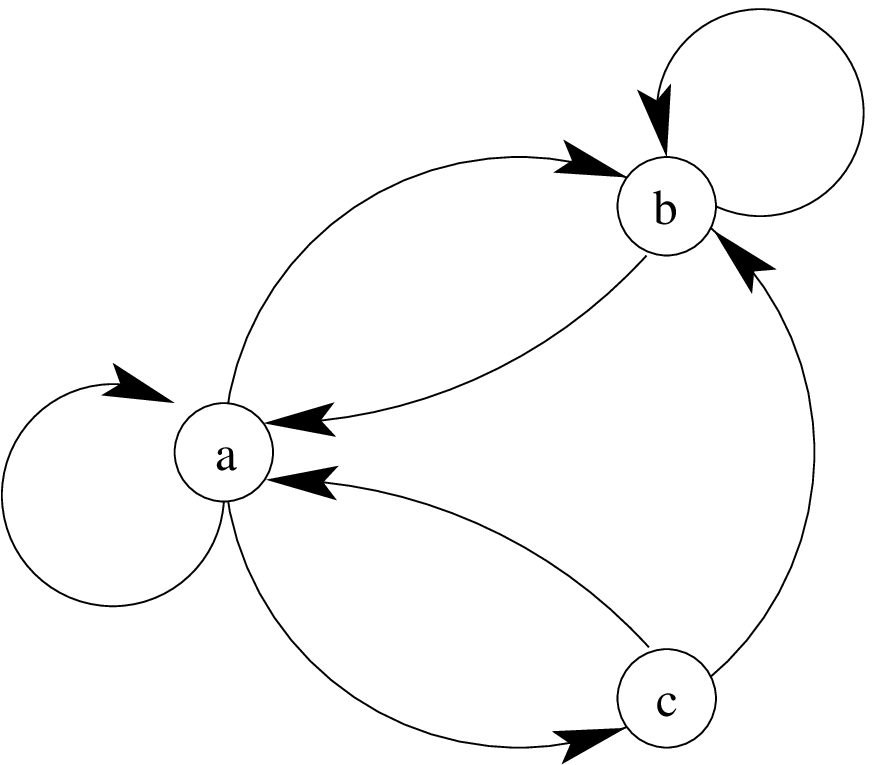}
$$
\end{minipage}\hfill
\begin{minipage}{5cm}
$$
C_1=\left (
\begin{matrix}
0 & 0 & 0\cr
0 & 0 & 0\cr
0 & 1 & 0\cr
\end{matrix}\right )
$$
\end{minipage}\hfill
\begin{minipage}{5cm}
$$
C_2=\left (
\begin{matrix}
0.9 & 1.2 & 0.9\cr
0.9 & 1.2 & 1\cr
0.9 & 1.2 & 1\cr
\end{matrix}\right )
$$
\end{minipage}

We consider two distinct quantities $q_1$ and $q_2$. Their associated cost matrices are respectively $C_1$ and $C_2$. $q_1$ counts the number of times that transition $c\rightarrow b$ is taken. $q_2$ models the concentration of a product. It increases by $20\%$ for every transitions pointing to $b$  and decreases by $10\%$ for all other transitions (\textit{i.e.,} abstraction of the product natural degradation). 
 Note here that, by convention, a cost of 0 (or respectively 1 for multiplicative quantities) has been assigned to the non existing transitions $b\rightarrow c$ and $c\rightarrow c$. 
Thus, as illustration, given initial quantities $q_1=0$, $q_2=1$ and a trajectory, \textbf{abacbacacba}, their values become $q_1=2$ and $q_2=0.826$.

\subsection{Probabilistic model}

On applicative purpose, we are interested in the values of all quantities for a random trajectory. Transitions impact differently to the global system behavior. We assume here that each transition possesses its own probability. Thus, at each step, one chooses randomly between all the transitions that leave the current state. The sum of the probabilities associated with all edges that leave a given state is 1.

For fixed probabilities at all steps, this model is a weighted Markov chain. Nevertheless, probabilities may vary, showing a behavior controlled by a dynamical system  (see~\cite{vallee} for a further details about dynamical sources). This model is hence quite general and particularly accurate for theoretical studies since it includes simple probabilistic models such as Markov chains, Hidden Markov chains or trickiest models that handle unbounded correlations (\textit{i.e.,} the choice made at one step influes on all next choices). In this last case, generating operators play the role of transition probabilities. For this reason, we assume our model as a \emph{graph with dynamical sources} (or GDS model). 
 The GDS is called \emph{nice} if it satisfies some classical conditions of the theory of Markov chains and dynamical sources (namely, the graph is strongly connected and aperiodic and all the dynamical systems are topologically mixing and possess expansive branches).  We consider the transition matrix $T=(t_{i,j})$ of the qualitative graph in which the element $(i,j)$ is the generating operator relative to the transition from $i$ to $j$. Reasoning on system properties implies to focus on quantities asymptotic properties. These mathematical properties are well studied in both theories of Markov chains and dynamical sources~\cite{Bourdon:2006}.

\subsection{Typical behaviors}
Previous theoretical assumptions allow us to emphasize  typical characteristics of quantities. More precisely, for a given GDS model,  we provide results for the mean, the variance and the limit distribution. The following theorem synthesizes our results.

\begin{theorem}
Let $\mathcal{M}$ by a nice GDS model with transition matrix $T$ and $q$ a quantity with cost matrix $C$. Let $Q_n$ be the random variable equal to the quantity $q$ after $n$ steps of the GDS model $\mathcal{M}$.
\begin{enumerate}
\item[$(i)$]{if $q$ is an additive quantity, $Q_n$ follows asymptotically (when $n$ tends to $\infty$) a Normal law with mean and variance
$$
\esp{Q_n}=\alpha_1 n +O(1)
\qquad
\var{Q_n}=\alpha_2 n + O(1),
$$
where $\alpha_1=\lambda'(1)$ and $\alpha_2=\lambda''(1)+\lambda'(1)-\lambda''(1)^2$ express by means of derivatives of the dominant eigenvalue of the matrix $\mathbb{A}(u)$ defined by $\mathbb{A}_{i,j}(u)=T_{i,j} u^{C_{i,j}}$.
}
\item[$(ii)$]{if $q$ is a multiplicative quantity,  $Q_n$ follows asymptotically (when $n$ tends to $\infty$) a $\log-$Normal law with mean and variance
$$
\esp{Q_n}=\beta_1\ \gamma_1^n +o(\Lambda_1^n)
\qquad
\var{Q_n}=\beta_2\ \gamma_2^n + o(\Lambda_2^n),
$$
where $\gamma_1 = \lambda (e)$ and $\gamma_2 = \max (\lambda(e^2), \lambda(e)^2)$ express by means of the dominant eigenvalue of the matrix $\mathbb{A}(u)$ defined by $\mathbb{A}_{i,j}(u)=T_{i,j} u^{\ln C_{i,j}}$. 
 $\beta_1$ and $\beta_2$   are constants corresponding to the dominant eigenvectors of $\mathbb{A}(e)$ and $\mathbb{A}(e^2)$. The error terms $\Lambda_1$ and $\Lambda_2$ verify $\Lambda_1 < \gamma_1$ and $\Lambda_2 < \gamma_2$.
}
\end{enumerate}
\end{theorem}

\begin{skproof}
See \cite{bourdon-2007} for a complete proof of this theorem. Considering the additive case is sufficient,
if $q$ is a multiplicative quantity, then $\log q$ is an additive one and it is easy to obtain the results of $(ii)$. The study involves several classical elements on the average-case analysis theory such as generating functions. 
Let $m$ be the number of states of the GDS model and $P_0$ be a probability vector whose element $i$ is the probability that initial state is state $i$. Since we consider asymptotic cases, this initial vector does not have any influence on the result. The generating function $Q(z,u)$ defined as
$$
Q(z,u) = \sum_{n\geq 0} P_0 z^n \mathbb{A}(u)^n \left ( \begin{matrix}1\\ 1\\ \vdots \\ 1\end{matrix} \right),
$$
permits to study the quantities of interest. Indeed, 
$$
\esp{Q_n} = \frac{\partial}{\partial u} [z^n] Q(z,u) |_{u=1}.
$$
For a nice GDS model, the matrix $\mathbb{A}(u)$ admits a dominant eigenvalue in a neighbourhood of $u=1$ and decomposes as $\mathbb{A}(u)=\lambda(u) \mathbb{P}(u) + \mathbb{N}(u)$, where $\lambda(u)$ is the dominant eigenvalue, $\mathbb{P}(u)$ is the dominant eigenvector and $\mathbb{N}(u)$ is associated to the remainder of the spectrum (and is thus orthogonal to $\mathbb{P}(u)$). Consequently, for large $n$, one has
$$
\mathbb{A}(u)^n\approx \lambda(u)^n \mathbb{P}(u).
$$
It is easy to obtain a formula for the mean. The study of the variance follows similar assumptions and involves the second derivative of $Q(z,u)$. Finally, the limit law is obtained by applying Hwang's \cite{Hwang:1996} general result on bivariate generating functions.
\end{skproof}

\medskip
Supplementary results have been obtained but they are not detailed here. Among others, we calculate the probability for a quantity to attain a given threshold before a given time $t$ (it generalizes the hitting probability, common in the Markov chain theory) and the joint law of several quantities. Most on our computations extends in same cases when the graph is not strongly connected or aperiodic.

\subsection{A typical biological study}
Previous theories allow us to reason on system quantitative properties but provide as well the core of a dedicated software\footnote{POGG: Probabilities On Genetic Graphs is available at \url{http://www.sciences.univ-nantes.fr/lina/bioserv/POGG/}}. This software works on GDS models with fixed probabilities and represents an accurate tool for simulating macro-molecular networks. As inputs, it needs a graph (or a qualitative graph in a better case)  and  the cost matrix of quantities of interest. GDS probabilities are unknown or partially unknown which make almost impossible to predict the quantitative impact of an interaction on the system behavior. Nevertheless, experimental results of quantity behaviors, like protein concentrations, are known. 
POGG uses such an information and adopts a  reverse engineering point of view. Previous theoretical results give some (in)equalities that relate unknown significance probabilities to experimental measures (of part or all of the quantities). 
POGG uses general techniques of local search theory, such as Tabu search, for estimating the impact of a local interaction in the whole biological system. The determined model gives us the opportunity to predict the behavior of others quantities.  Note that the software also provides supplementary informations such as an approximation of the hyper-volume of models that are consistent with the measures. This information helps to decide whether a new measure is informative or not, using a simple comparison between different volumes.

\section{Results}
\cite{Ropers:2006aa} models the growth phase transition of a bacteria after a nutritional stress.  In particular, the model shows  the abandon of exponential growth state to a more stationary growth during a carbon starvation stage. Their qualitative results are relevant with experimental knowledges, which allows us to consider the model as an appropriate benchmark for our modeling approach. Furthermore, macromolecules that interact within the model are well studied. It gives us various partial quantitative informations that have to be introduced into the qualitative model. 

\subsection{Carbon starvation response in \textit{Escherichia coli}: gene regulatory network and qualitative rules validation}
\begin{figure}[h]
\begin{center}
\includegraphics[scale=.6]{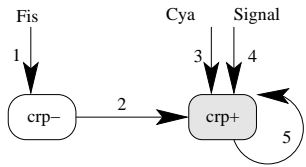}
\caption{Qualitative representation of \textit{crp} interaction with others genes and carbon starvation signal. 1 represents the repression of \textit{crp} by Fis protein production. 2 and 5 are transitions for the basal synthesis rate which plays an important role during the exponential growth phase. Combination of concurrent rules 3 and 4 synthesizes the \textit{crp} activation via cAMP metabolite.}
\label{crp}
\vspace{2 mm}
\end{center}
\end{figure}

We consider similar hypotheses to those exposed in \cite{Ropers:2006aa} and propose a new graph that represents identical qualitative behaviors of bacterial responses after a nutritional stress. For illustration and using abstractions described in Sec.~\ref{formalization}, we detail in Figure~\ref{crp}  one particular biological component: \textit{crp} gene. The gene \textit{crp} is controlled by two promoters that are both repressed by Fis protein  \cite{GonzalezGil:1998aa}. Following assumptions from \cite{Ropers:2006aa}, we omit the negative control of \textit{crp} and summarize the impact of cAMP metabolite using rules that imply Cya and Crp protein and carbon starvation signal as well \cite{Harman:2001aa}. 

\begin{figure}[t]
\begin{center}
\includegraphics[scale=.6]{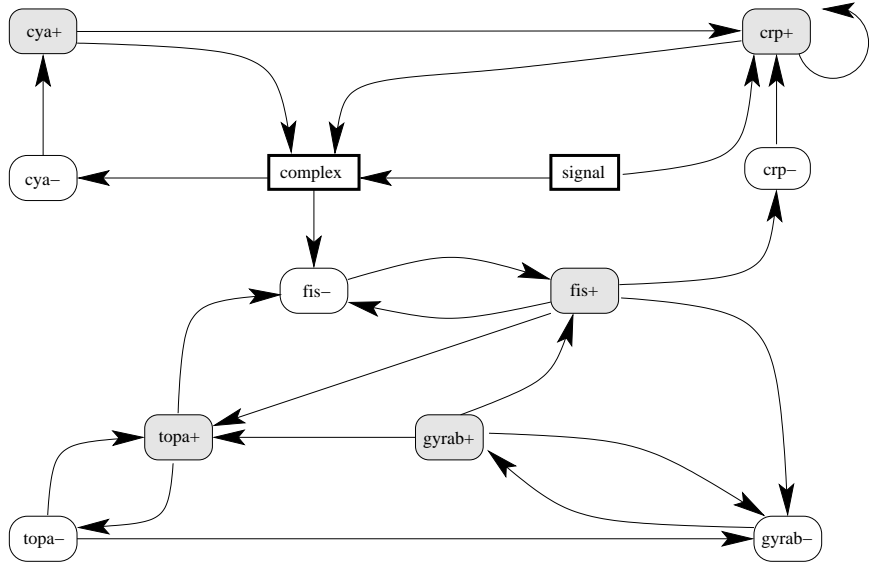}
\caption{Qualitative graph representing genes regulatory network of  carbon starvation response in \textit{Escherichia coli}. Signal represents the input module that indicates carbon starvation condition. }
\vspace{2 mm}
\label{ecoli}
\end{center}
\end{figure}

We use a similar approach for describing each biological component of the gene regulatory network. Figure~\ref{ecoli} represents the corresponding qualitative graph. Our aim is to demonstrate advantages of our probability approach. Therefore we will not detail here biological assumptions that have been used for building the model. See \cite{Ropers:2006aa} for exhaustive hypotheses. 
Before further in silico investigations, the model has to be validated. It is the \textit{sine qua non} condition for applying a probabilistic approach. Although probabilities can be estimated using an appropriate optimization, our confidence in such parameters is related to the ability of the model to reproduce appropriate qualitative behaviors of the biological system (\textit{i.e.,} various kinds of qualitative models can produce similar quantitative results). 
Using the symbolic model-checker of BIOCHAM \cite{Calzone:2006aa,FageSoliChab04}, we thus check qualitative rules in order to verify their consistency with experimental understandings. In particular, \cite{Browning:2004aa} shows an antagonistic relationship between \textit{fis} and \textit{crp} activities.
For validating the model, we are able to ask positive queries (\textit{i.e.} queries where the expecting answer is true) such as 
$(\textit{fis}^+ \wedge \neg \textit{fis}^- ) \implies ( \textit{crp}^- \wedge \neg \textit{crp}^+)$. 
We are as well able to ask negative queries (i.e. queries where the expecting answer is false), such as $(\textit{fis}^+ \wedge \neg \textit{fis}^- ) \equiv (\textit{crp}^- \wedge \neg \textit{crp}^+)$. Using this formal verification on the qualitative model, we successfully check other biological properties  like the relationship between the carbon starvation signal and \textit{crp} expression \cite{Ishizuka:1994aa} as well with  \textit{cya} activities \cite{Ball:1992aa}. 

\subsection{Probabilistic results}
\begin{figure}[t]
\begin{center}
\includegraphics[scale=.62]{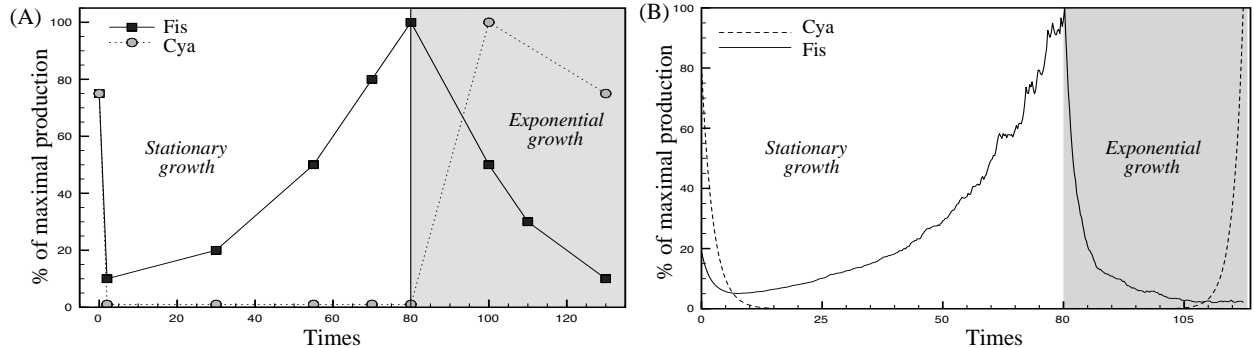}
\caption{Summary of informations used and produced by the probabilistic approach. (A) shows variations of Fis and Cya protein concentrations in function of growth phases \cite{Ball:1992aa,data1}. Two Fis variations during stationary growth have been used for estimating probabilities associated with qualitative transitions from Figure~\ref{ecoli}. It allows to reproduce quantitative behaviors of Cya and Fis during both growth phases (B). 
}\vspace{2 mm}
\label{result}
\end{center}
\end{figure}

Therefore, we have at our disposal an accurate qualitative graph (Figures~\ref{ecoli}) and quantitative informations (Figure~\ref{result}~(A)) that belong to the same bacterial system. Our modeling approach exploits such informations and predicts probabilities on graph transitions using a local search algorithm. In practice, we take into account the fact that Fis concentration is multiplied by 10 in 80 minutes during the stationary growth phase. We assume Fis concentration $q_{\textit{Fis}}$  as a multiplicative quantity (see Sec. 2.2). Therefore it increases by $20\%$ for each transition pointing to $\textit{fis}^+$, decreases by $20\%$ for all transitions pointing to $\textit{fis}^-$ and decreases by $10\%$ for natural degradation passing through all other transitions.  We estimate the Fis quantity at time 80, $q_{\textit{Fis}}\approx 10 \cdot \gamma^{1600}$ with $\gamma \approx 1.001$ (1600 corresponding to the number of steps performed by the model during 80 minutes, this number is established by considering Cya natural degradation during the first 2 minutes). Comparing this numerical value with constants from Theorem 1, we get a constraint that relates probabilities with a measure on the system. Local search methods allow to find a suitable probability matrix used for simulations.

Figure~\ref{result}~(B) shows the estimated variation of Cya and Fis protein in function of growth phases. During the stationary growth, our model accurately predicts a decrease followed by an increase of Fis protein production \cite{Ball:1992aa}.  It emphasizes the ability of our approach to spread partial quantitative knowledges through the qualitative network. Despite a quantitative estimation using two measures during the stationary phase, interestingly, our model predicts efficiently the Fis concentration decrease during the exponential phase. This model artifact represents a quantitative emerging property of the biological system which gives insights about global behaviors.
 
 Estimative Cya protein variations are as well consistent with experiments during stationary phase. However, despite an appropriate increase during the beginning of the exponential phase, the Cya production does not follow an expected peak \cite{data1}. It mights reflect a shortcoming or a missing qualitative transition that represses the \textit{cya} gene. We consider such an information as a guidance for future models or further experiments that might focus on \textit{cya} gene regulations. 

A close attention to estimated probabilities gives results that are related with the quantitative sensitivity of the model. More precisely, an estimation of the hyper-volume associated with the model emphasizes whether a new measure is informative or not. Our model shows that the probability associated with \textit{topa$^+$} and \textit{fis$^-$} transition is highly constrained in order to maintain an overall consistency between heterogeneous informations. This transition is a shortcut adapted from \cite{Ropers:2006aa} for representing DNA supercoiling effect on \textit{fis} gene expression. Experiments suggest that \textit{fis} is involved in fine tuning of the homeostatic control of DNA supercoiling \cite{Schneider:2000aa}.  A small change in DNA supercoiling drastically affects the \textit{fis} expression. This information is accurate with our estimative impact of this transition on the global system behavior.

\section{Discussion}

Recent fruitful probabilistic approaches has been developed for studying gene regulatory networks \cite{pnet1, pnet2, pnet3}. These approaches add probabilities to an already defined deterministic model. It gives the opportunity to study probability variation impacts and eventually to determine probability sets that accurately represent experiments. 
%More precisely, the aim of PBN is to obtain the steady-state distribution of the state graph of the system (here, for $n$ genes, the state graph possesses $2^n$ vertices and the computation of the stationary distribution is tricky). For Bayesian networks, 
%observations are related by conditionnal probabilities with their possible causes, and as a consequence, it is possible to determine the most probable cause of a given observation. T
%the point is to infer valid conditional probabilities between causes and their effects in order to determine the most probable cause of a set of observed effects.
Knowing the transition probability graph, the major issue of these approaches is to compute the asymptotic (stationary) distribution and to reason on it. 

Our original method appears as a complementary approach that adds new natural informations in a general probabilistic graph. It gives the opportunity to reason on emerging system properties by focusing on asymptotic properties of the probabilistic model. We prove that their asymptotics are related to natural constants on a weighted transition matrix. The proposed method allows to design constraints between probabilities and observations, which gives the opportunity  to deal with unkwown transition probabilities. Therefore our results are adapted to a large class of probabilistic models and their integration within a more general framework such as PBN and Bayesian networks seems promising.

The number of biological details at disposal defines the model abstraction level which conduces to choose an accurate biological abstraction. It is more or less discrete in function of the number of qualitative states. Our probabilistic-like technique is able to combine quantitative informations with various qualitative abstractions of biological systems, \textit{i.e.,} from boolean to PDE network  \cite{deJong:2002aa}. Therefore, our method emphasizes a convenient flexibility for analyzing biological systems because it presents  major advantages for integrating heterogeneous knowledges such as those that constitute the \textit{Escherichia coli} starvation system.

During this study, various biological models were elaborated. After probability optimization, most of them give relevant quantitative simulation results. Nevertheless, they remain inconsistent with their ability to reproduce the whole set of expecting experimental behaviors. It hence confirms the support of reasoning  rather than just similating that prevents to validate the model using few simulations.  Furthermore, it emphasizes the need for an appropriate qualitative validation of model behaviors prior to apply our probabilistic technique. In this purpose, the biological system has been described using a set of original qualitative rules. It allows us to use a formal verification technique in a qualitative validation requirement. Therefore our technique appears as a natural extension of regular qualitative modeling approaches for extending robust qualitative models toward quantitative properties. 

\begin{small}
\bibliography{bourdon-etal}
\bibliographystyle{genres}
\end{small}
\end{document}